\documentclass[nofootinbib,superscriptaddress,twocolumn]{revtex4-1}

\usepackage{amsmath}
\usepackage{amsfonts}
\usepackage{times}
\usepackage{graphicx}

\begin{document}

\title{OPERA-reassessing data on the
  energy dependence of the speed of neutrinos}

\author{Giovanni AMELINO-CAMELIA}
\affiliation{Dipartimento di Fisica, Universit\`a di Roma ``La Sapienza", P.le A. Moro 2, 00185 Roma, EU}
\affiliation{INFN, Sez.~Roma1, P.le A. Moro 2, 00185 Roma, EU}

\author{Giulia GUBITOSI}
\affiliation{Department of Physics, University of California, Berkeley,
CA 94720, USA}

\author{Niccol\'o~LORET}
\affiliation{Dipartimento di Fisica, Universit\`a di Roma ``La Sapienza", P.le A. Moro 2, 00185 Roma, EU}
\affiliation{INFN, Sez.~Roma1, P.le A. Moro 2, 00185 Roma, EU}

\author{Flavio MERCATI}
\affiliation{Departamento de F\'isica T\'eorica, Universidad de Zaragoza, Zaragoza 50009, Spain}

\author{Giacomo ROSATI}
\affiliation{Dipartimento di Fisica, Universit\`a di Roma ``La Sapienza", P.le A. Moro 2, 00185 Roma, EU}
\affiliation{INFN, Sez.~Roma1, P.le A. Moro 2, 00185 Roma, EU}

\author{Paolo LIPARI}
\affiliation{INFN, Sez.~Roma1, P.le A. Moro 2, 00185 Roma, EU}

\begin{abstract}
\noindent
We offer a preliminary exploration of the two sides of the challenge provided
by the recent OPERA data on superluminal neutrinos. On one side we stress that some aspects
of this result are puzzling even from the perspective of the wild quantum-gravity literature,
where arguments in favor of the possibility of superluminal propagation have been presented,
but not considering the possibility
of such a sizeable effect for neutrinos of such low energies.
We feel this must encourage particularly severe scrutiny of the OPERA result.
On the other side, we notice that the OPERA result is reasonably consistent
with $\mu$-neutrino-speed data previously obtained at FERMILAB, reported in papers of 2007 and 1979.
And it is intriguing that these FERMILAB79 and FERMILAB07 results, when combined
 with the new OPERA
result, in principle provide a window on $\mu$-neutrino speeds at different energies
broad enough to compare alternative phenomenological models.
We test the discriminating power of such an approach
by using as illustrative examples
the case of special-relativistic tachyons, the case of ``Coleman-Glashow-type"
momentum-independent violations of the special-relativistic speed law,
and the cases of linear and quadratic energy dependence of the speed of ultrarelativistic
muon neutrinos. Even just using $\mu$-neutrino data
 in the range from  $\sim$ 3 GeVs to $\sim$ 200 GeVs the special-relativistic tachyon
 and the quadratic-dependence case are clearly disfavoured. The linear-dependence case gives
a marginally consistent picture and the Coleman-Glashow scenario fits robustly the data.
We also comment on Supernova 1987a and its relevance for consideration of other neutrino species,
also in relation with some scenarios that appeared in the large-extra-dimension literature.
\end{abstract}

\maketitle

\baselineskip11pt plus .5pt minus .5pt

\section{Preliminaries and motivation}
The OPERA collaboration recently reported~\cite{opera}
evidence
of superluminal behavior for $\mu$ neutrinos ($\nu_\mu$).
Specifically, these ``OPERA/CNGS11 data"
(data reported by the OPERA collaboration, CNGS beam, in 2011~\cite{opera})
imply for the speed of such neutrinos the estimate
\begin{eqnarray}
{v}_{\nu_\mu} - 1= (2.48 \pm 0.28~(stat)~ \pm 0.30~ (sys)) \cdot 10^{-5},
\label{mammamia}
\end{eqnarray}
a significance of six standard deviations (we use units such that the speed of light is $c=1$).

This would be the most significant discovery in fundamental physics over the last
several decades, so the OPERA data will inevitably motivate
a healthy exploration of both possible outcomes:
on one side the data, particularly the possibility of unnoticed systematic biases,
should be scrutinized very carefully, and on the other side,
which however will require (also in light of some of the considerations we
here offer) the guidance of some dedicated model building,
one may look for corroborating evidence in totally independent measurements,
at different energies using different techniques.

We here report a preliminary exploration of these two sides.\\
We are going to provide further motivation for careful scrutiny of the
data by observing that such a result cannot be accommodated in any reasonably well
studied existing theory speculation. In particular, even in
the quantum-gravity literature,
small parts of which have provided motivation for searches of violations
of Lorentz symmetry, including some proposals of superluminal type,
one finds no scenario with an effect of that magnitude for particles with
energies so far from the Planck scale.

For the opposite side, the one of attempts to find
evidence corroborating the OPERA/CNGS11 result,
also exploiting the guidance of some dedicated phenomenological models,
our main message is that combining CNGS11 with other previously-obtained data
on the speed of $\mu$-neutrinos we might have sufficient guidance
to filter significantly the list of
candidate phenomenological models.
We find that a particularly interesting picture is obtained when combining
the CNGS11 data with the ``FERMILAB07" data (obtained at Fermilab,
by the MINOS collaboration, and reported in 2007~\cite{minos2007})
and the ``FERMILAB79" data (also obtained at Fermilab,
and reported in 1979~\cite{preminos1979}). This allows us to look at
 speed of $\mu$-neutrinos with data populating with acceptable density
the range from $\sim 3$GeV to $\sim 200$GeV.
So if one takes these data at face value (as a working assumption, looking for
evidence possibly corroborating CGNS11) one has a criterion to select
phenomenological models, whose guidance could be used to set up
particularly meaningful other tests of the superluminal $\mu$-neutrino hypothesis.

Mostly as a way to test that our proposal of combining the new CGNS11 data
with the previous FERMILAB07 and FERMILAB79 data actually can ``discriminate models"
(though only conditionally on the working assumption that the data can be taken
at face value), we focus on a few very simple
phenomenological pictures of superluminal
particles: a ``standard" (imaginary-mass)  special-relativistic tachyon,
the case of ``Coleman-Glashow neutrinos"~\cite{colgla1,colgla2},
with real mass and violatons of the special-relativistic
speed law which are momentum independent in the ultrarelativistic regime,
and two ``DSR-type pictures"~\cite{dsrnature,gacdsrIJMPD2002,gacdsrPLB2001,kowadsr,leedsrPRL},
with real mass, possibly ``deformations" (rather than preferred-frame breakdowns)
of Lorentz symmetry, and momentum-dependent speed in the ultrarelativistic regime.
We expose some discriminating power for the strategy here proposed by finding
that, among these illustrative examples of phenomenological models,
the special-relativistic-tachyon scenario and the DSR model with quadratic dependence
on the deformation scale are clearly disfavored by
CGNS11+FERMILAB07+FERMILAB79 data, while, at least adopting a conservative approach,
 the Coleman-Glashow scenario and the DSR model
with linear dependence give a reasonably good description.
We also comment on Supernova 1987a and its relevance for consideration of other neutrino species,
also in relation with some scenarios that appeared in the large-extra-dimension literature.

 \section{No quantum-gravity models, but some simple phenomenological test theories}
The physics literature provides
 no compelling argument for special-relativistic tachyons
 and only some rare instances where
 departures from standard Lorentz symmetry
 could be motivated.
 We feel that the most compelling arguments for possible departures
 from standard Lorentz symmetry are found in the part of the quantum gravity
 literature which motivates~\cite{grbgac,gampul,mexweave,leeQGLIV}
 the adoption of a nonclassical-geometry description of spacetime,
 with associated violations or deformations of Lorentz symmetry.
 Moving from the level of ``theories" to the one of ``phenomenological pictures"
 a noteworthy possibility is the much-studied idea of
 large extra dimensions, within which several authors
 have motivated mechanisms for violations of Lorentz symmetry
 (see, {\it e.g.}, Refs.~\cite{ammosovvolkov,volkov,strumia,DentPas,HollePas} and references therein).
 	
 We actually postpone a detailed discussion of the relevant quantum-gravity
 literature to a forthcoming companion paper~\cite{companion}, since it would
 impose here a lengthy aside.
 We do note here that
 superluminal particles have been motivated in parts of the quantum-gravity
 literature, and aspects of the quantum-gravity problem offer solid motivation
 for rather strong particle dependence of the effects (so that it would
 not be surprising to find the superluminal behavior of neutrinos, possibly even just some types of
 neutrinos,
 to be a few orders of magnitude stronger than for other particles).
 But the bottom line is that~\cite{companion} the effect reported by OPERA/CGNS11
 appears to be much too strong to be of quantum-gravity origin:
 one could rather compellingly motivate from quantum gravity the qualitative
 nature of that experimental result but the magnitude appears to be
 unbelievably gigantic
 by quantum-gravity standards.

 The OPERA result is in the peculiar position not only of conflicting with
 one of the cornerstones of current theories, but also of conflicting with
 the most appealing alternatives.
We are fully aware
 of the fact that some major discoveries in the history of physics
 shared this fate, but nonetheless we feel this must be viewed as an invitation
 to be cautious with theory speculations inspired by the OPERA result,
 and also as an invitation to scrutinize the result meticulously.

 Going back to the issue of the puzzling magnitude of the OPERA effect,
 as seen from a quantum-gravity perspective,
let us here be satisfied to discuss it
 specifically within (two versions of)
 one of the pictures considered in the quantum-gravity
 literature, a picture from which we shall also borrow inspiration for
 two of the test theories here used as illustrative examples
 of ``pure-phenomenology interest".
 Let us start this off with the dispersion/on-shellness relation
\begin{equation}
E^2 = m^2 + p^2 + \ell_1 E p^2~,
\label{led}
\end{equation}
which
 has been much studied in parts of the quantum-gravity literature.
 It is a scheme with particles of ordinary real mass ($m^2 >0$),
 but it is  superluminal, as easily verified computing
 the speed $dE/dp$ in the ultrarelativistic regime $E \gg m$ of (\ref{led}) (for positive $\ell_1$).
 Evidently such a dispersion relation cannot be accommodated within standard
 special-relativistic Lorentz symmetry, but interestingly it is possible to implement
 this dispersion relation in a deformed-Lorentz-symmetry framework,
 the so called ``DSR" framework~\cite{dsrnature,gacdsrIJMPD2002,gacdsrPLB2001,kowadsr,leedsrPRL},
 without any breakdown of the relativity of reference frames: in such DSR frameworks
 the laws of transformation between inertial observers are $\ell_1$-modified,
 but the equivalence of all inertial observers is preserved.
 A few hundred papers have been devoted to this possibility, within
 the quantum-gravity literature, over the last decade,
 but all of them assume that the deformation scale $\ell_1$ should be
 of the order of the ``Planck length", {\it i.e.} $(10^{19}GeV)^{-1}$,
 or at best (and only if confined to spin-$1/2$ particles)
 a few orders of magnitude greater than that.
 The Planck length is the only natural scale of the quantum gravity problem,
 but if we use (\ref{led}) with $\ell_1$ of the order of the Planck length
 then the conclusion would be for OPERA neutrinos to be affected only
 at the level of 1 part in $10^{18}$, rather than 1 part in $10^{5}$:
 the qualitative feature is not completely foreign to the quantum-gravity literature
 but the magnitude of the effect is off by 13 orders of magnitude.
 Another way to characterize the situation equivalently is to state that the OPERA
 data appear to probe, from the viewpoint of (\ref{led}),
 a scale of $\ell_1 \sim (10^{6}GeV)^{-1}$, which is a small scale by the standards
 of other areas of physics but is a gigantically macroscopic length scale by the standards
 of quantum-gravity researchers.

One easily sees that this point applies also
 to the similar case of a quadratic formula
\begin{equation}
E^2 = m^2 + p^2 + \ell_2^2 E^2 p^2
\label{ledquad}
\end{equation}
which also has been at the center of strong interest in parts of the recent
quantum-gravity literature.
For this sort of quadratic corrections the OPERA/CNGS11 neutrinos would be probing
the even more gigantic length scale of $\ell_2 \sim (10^{3}GeV)^{-1}$, 16 orders of magnitude
bigger than the Planck length!!

 We shall not dwell on this any further here. We shall be satisfied to have contributed
 a bit to the first assessments of how surprising this result must be considered.
 We feel it would be legitimate to argue that the OPERA result is so puzzling that
 it should be completely ignored until it somehow gets confirmed independently.
 But, as anticipated above,
 we here take the attitude that the OPERA result itself, when combined with
 other available results on the speed of $\mu$ neutrinos, can provide guidance
 for attempts of confirming its result truly independently, in other energy regimes,
 using different experimental techniques. And we expect that
 the development of suitable test theories could play a crucial role for guiding
the setup of any such attempts.

 We illustrate here how we envisage this interplay between available data
 and test theories, by considering a few rudimentary test theories of
 the energy dependence of the speed of $\mu$ neutrinos
 and assessing on them the discriminating power
 of our approach.\\
 Our first candidate is the ``standard" special-relativistic tachyon, which
 in the ultrarelativistic regime has speed
\begin{equation}
v = \sqrt{1+\frac{{\cal M}^2}{E^2}} \simeq 1 + \frac{1}{2} \frac{{\cal M}^2}{E^2} \label{tachy}
\end{equation}
where for convenience we introduced the parameter ${\cal M}^2$ which
is {\underline{positive}} (the opposite, ${\cal M}^2 = - m^2$, of the square of the imaginary
mass $m$ of a special-relativistic tachyon).

Our second candidate is inspired by work of Coleman and Glashow~\cite{colgla1,colgla2}
which would lead to the following description of the dependence of speed on energy
in the ultrarelativistic regime
\begin{equation}
v \simeq 1- \frac{1}{2} \frac{m^{2}}{E^{2}} + \delta \label{colgla}
\end{equation}
where $\delta$ is the parameter characterizing the maximum attainable speed
by the particle. Evidently this maximum attainable speed may be greater than $c$,
if $\delta$ is positive, and yet also (\ref{colgla})
assumes~\cite{colgla1,colgla2} the mass $m$ to be real.
This Coleman-Glashow picture evidently requires departures from the special-relativistic
description of Lorentz symmetry. And there are no established results
on possibly implementing this
as a ``deformation" of Lorentz symmetry, so it has been studied exclusively as
a scenario for a full ``breakdown" of Lorentz symmetry, including the emergence of
a preferred frame. But it is a much studied framework, which has shown some robustness
(for what concerns logical consistency) in several attempted applications, and can be
even accommodated naturally within the more general
framework of ``Standard Model Extension"~\cite{sme1,smerev}.

For our third  and fourth
candidates we go back to the DSR-type scenarios already briefly
mentioned above, for Eqs.~(\ref{led}) and ~(\ref{ledquad}). Evidently the OPERA/CNGS11 result could
not fit within the original spirit of studies of these scenarios, since as mentioned
the effects are much larger than expected. But, as we are looking for candidates to
test the ``phenomenology content" of data so far available on $\mu$ neutrinos,
we felt having some DSR-type candidates could be valuable, at least for illustrating
the idea of departures from special relativity which do not require introducing
a preferred frame. So for the third scenario we take, from (\ref{led}),
the following description of the dependence of speed on energy
in the ultrarelativistic regime\footnote{The same formula
for the  dependence of speed on energy can of course also be introduced~\cite{grbgac,ellisMAGIC,unoEdue}
in scenarios where Lorentz symmetry is fully broken, with a preferred frame.
Actually the first study that brought this proposal to the attention
of the quantum-gravity community, the one in Ref.~\cite{grbgac}, assumed broken Lorentz symmetry,
and only later it was realized that a DSR-type formulation was also
possible~\cite{dsrnature,gacdsrIJMPD2002}.
We here label these formulas as ``DSR-type" because it is relevant for the
(however limited) representativity of the small sample of scenarios we consider that
two of the scenarios are also known to be compatible with the Principle of Relativity of inertial
frames (``only" at the cost of adopting DSR-type~\cite{dsrnature,gacdsrIJMPD2002} deformed laws
of transformation between inertial observers).}
\begin{equation}
v \simeq 1 - \frac{1}{2} \frac{m^{2}}{E^{2}} + \ell_1 E \label{dsr}
\end{equation}
where $\ell_1$ is the parameter characterizing the dependence (in this case linear)
of the speed of ultrarelativistic particles on momentum. While evidently
the proposal of this DSR scheme~\cite{dsrnature,gacdsrIJMPD2002} had in mind that $\ell_1$ would
be at least roughly of the order of the Planck length, we shall here allow it, for the
sake of the argument, to be even much greater than that.

Similarly we take inspiration from the quantum-gravity interest in
Eq.~(\ref{ledquad}), to consider here (once again far away from the originally intended,
Planckian, range of scales)
the phenomenology of
\begin{equation}
v \simeq 1 - \frac{1}{2} \frac{m^{2}}{E^{2}} + \ell_2^2 E^2 \label{dsrquad}
\end{equation}
where the parameter $\ell_2$ evidently charaterizes the quadratic dependence
of the speed of ultrarelativistic particles on momentum.

A noteworthy observation is that the large-extra-dimension literature has provided
some arguments motivating a superluminal behaviour of neutrinos,
but confined to some range of scales. Staring at the OPERA result it is in particular
striking to look back at papers such as Refs.~\cite{ammosovvolkov,volkov}, which
more than a decade ago, argued for superluminal behaviour of neutrinos
with onset at a scale not far from the electroweak scale.
And more recently (but much before the OPERA study) other large-extra-dimension
investigations have reported the possibility of superluminal behaviour
for neutrinos, confined to specific ranges of energy, as discussed in particular
 in Refs.~\cite{DentPas,HollePas} (also see Ref.~\cite{Ellis:2008fc}
for related phenomenology work in preparation for OPERA).
This literature appears not to provide us with a clear candidate test theory
that would be relevant here for our purposes,
but we shall take into account its indications when, in later parts of this manuscript,
we move away from our main focus on $\mu$ neutrinos in the $3-200$GeV range,
and consider briefly the behaviour of other species of neutrinos in other
ranges of energy.


\section{OPERA+FERMILAB07}
First reactions to the OPERA/CNGS11 result are of great astonishment.
From the theory perspective we fully ascribe, as stressed above, to that sort of reaction.
But on the other hand we feel there is too little awareness of the
fact that, if one looks exclusively
at previous experimental results on the speed
of $\mu$ neutrinos, this OPERA result does not really ``come out of the blue".
We shall here show that the CNGS11 result actually fits rather naturally
 with the ``FERMILAB07" data (obtained at Fermilab,
by the MINOS collaborations, and reported in 2007~\cite{minos2007})
and with the ``FERMILAB79" data (also obtained at Fermilab,
and reported in 1979~\cite{preminos1979}).

In this short section let us just compare the  OPERA/CNGS11 result
with the FERMILAB07 result reported in Ref.~\cite{minos2007}:
\begin{equation}
v -1= \left(5.1\pm 2.9 \right)\cdot  10^{-5}\label{minos}
\end{equation}
for $\mu$ neutrinos\footnote{Because of the exploratory goals
of our analysis we shall not dwell much on the details of the composition
of the data. If we label a result as a $\mu$-neutrino result it simply means
we expect the contamination from other neutrino species to be ``small".}
 of $\sim 3 GeV$.

This previous FERMILAB07 result (while intrinsically having only a significance
 of less than two standard deviations) goes in the same direction as the new
CNGS11 result, and the two results appear to be rather naturally compatible with each other.
Actually, they are not only compatible but they were also determined at relatively close energies,
and this  for our purposes is not welcome: if one only  combines
the OPERA/CNGS11 result and the FERMILAB07 result one gets very limited
ability to discriminate between models predicting different forms of energy dependence.


\section{OPERA+FERMILAB07+FERMILAB79}
It must be evident at this point that our main interest is in the
form of energy dependence of the speed of neutrinos (actually primarily $\mu$ neutrino, see later).
This is also reflected in the choice of illustrative examples of test theories
on which we focus: they are simple and representative of significant classes of
related speculations, and they also represent alternative
options for how the speed of ultrarelativistic neutrinos could depend on energy.
And it is also evident that because of these objectives it is desirable for
us to consider data at relatively high energies, at least somewhat higher
that the range covered by OPERA/CGNS11 and MINOS/FERMILAB07.
From this perspective it is interesting to reconsider the FERMILAB79 data
on  the speed of neutrinos reported in 1979 in Ref.~\cite{preminos1979}.
These extend all the way from $\sim 30$ GeV to $\sim 200$ GeV,
so they open a very valuable window for our purposes.

The energy dependence is so crucial for our purposes that we shall here not
use OPERA's most significant result, (\ref{mammamia}), obtained
by combining CC-internal and external events~\cite{opera}, since it carries no
verification of the neutrino energies. We find most valuable for our purposes to consider
only the CC-internal events for which~\cite{opera}
\begin{eqnarray}
({v}_{\nu_\mu}-1)\Big|_{14GeV}= (2.18 \pm 0.77~(stat)~ \pm 0.30~ (sys)) \cdot 10^{-5}
\nonumber\\
({v}_{\nu_\mu} -1)\Big|_{43GeV}= (2.75 \pm 0.75~(stat)~ \pm 0.30~ (sys)) \cdot 10^{-5}
\nonumber
\end{eqnarray}
These are of lower significance than (\ref{mammamia}), but carry the
energy information crucial for our purposes.

In our Fig.~1 we show the neutrino data\footnote{We here
do not consider the antineutrino data also found in Ref.~\cite{preminos1979}.
we feel in this exploratory stage it is an asset to look exclusively at $\mu$-neutrino data.} from Fig.~3
of Ref.~\cite{preminos1979}, together with the MINOS/FERMILAB07 result, and the OPERA
results we just noted.

\begin{figure}[htbp!]
\includegraphics[width=\columnwidth]{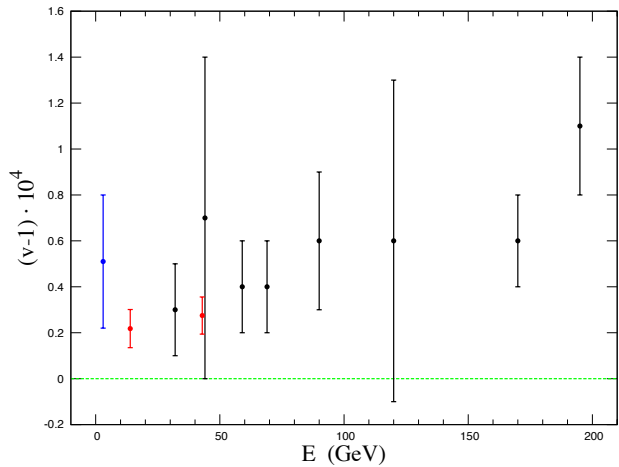}
\caption{The results for the speed of $\mu$ neutrinos reported by
OPERA (red), MINOS (blue) and in Fig.~3 of Ref.~\cite{preminos1979} (black).}
\end{figure}


Concerning the FERMILAB79 data we should stress that they actually concern
 the difference between the speed of the $\mu$-neutrino
and the speed of muons, but we shall be not too embarrassed of taking as working
assumption of this first exploratory study
 that the speed of muons, at least in that range of energies,
is faithfully described by standard special relativity, so we shall handle the FERMILAB79 data
as determinations of the speed of the $\mu$-neutrino at various neutrino energies.
More concerning for us is the fact that the analysis in Ref.~\cite{preminos1979}
relies very significantly on correcting for a large bias:
it was realized~\cite{preminos1979}
 that one should correct for the fact
that the relevant muons taking part in that differential measurement ended up being
on paths that were effectively longer than the path of the neutrinos they were ``racing" against.
The authors of Ref.~\cite{preminos1979} conclude that this would effectively produce a rigid
(equal at all energies)
downward shift of all estimates of the neutrino velocity, and that a very sizable such downward
shift should be applied.
Specifically this bias correction, which we shall denote with $b_{1979}$,
was estimated~\cite{preminos1979} at $b_{1979} = (0.5^{+0.2}_{-0.1}) \cdot 10^{-4}$.
Our Fig.~2 shows the effect of $b_{1979} = 0.5 \cdot 10^{-4}$ on the black points of Fig.~1.
As seen comparing Figs.~1 and 2 the nest result on the findings of Ref.~\cite{preminos1979}
roughly
 amounts to the subtraction of a large estimated ``background",
leaving the analysis with a small ``signal".

\begin{figure}[htbp!]
\includegraphics[width=\columnwidth]{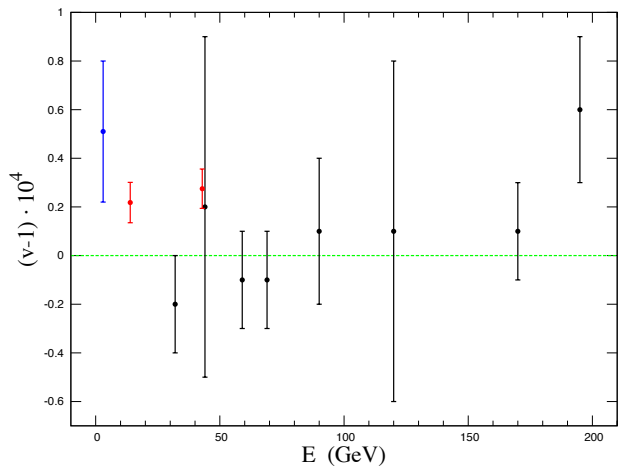}
\caption{As in fig.1 but taking into account the large bias correction
discussed at the end of Ref.~\cite{preminos1979}.}
\end{figure}

Looking at our Fig.~1 and even our Fig.~2 one cannot fail to notice that,
while surely previous measurements had not reached enough significance to make
substantial claims, the overall picture is perfectly consistent with what was then
very recently reported by OPERA.
And concerning the issue of a possible energy dependence
of the speed of ultrarelativistic muon neutrinos
we feel that it deserves at least some mention
that Ref.~\cite{preminos1979} reported an estimate of the dependence
on energy of the $\mu$ neutrinos, with linear law
and slope of $(0.3 \pm 0.1 ) \cdot 10^{-6}$, so an indication of non-vanishing slope
(which is visible in the summary of Ref.~\cite{preminos1979} data in our Figs.~1 and 2)
with a significance of three standard deviation. This went largely unnoticed but we feel
that in light of the results reported by OPERA it may require to be reconsidered.

One of our main points in this section is that the data of Ref.~\cite{preminos1979},
with information at different energies, may be a valuable resource for attempts
of interpretation of the OPERA result. We shall do this in the following, while
proceeding cautiously because of the concerns for the mentioned large bias estimate
given in Ref.~\cite{preminos1979} for their data.

\subsection{OPERA+FERMILAB07+FERMILAB79 and non-tachyonic special relativity}
It is easy to see that the comparison of standard non-tachyonic special relativity
with the black and blue (pre-OPERA) data points of our Fig.~1 could have produced some
serious concerns. But using the data of Ref.~\cite{preminos1979}, bringing in
the mentioned large bias correction, the situation is the one
of the black and blue points in Fig.~2,
which is reasonably consistent with
 non-tachyonic special relativity. The assumed pre-OPERA starting point
 was our Fig.~2, so non-tachyonic special relativity was considered to be in good health.

But non-tachyonic special relativity
definitely is no longer in good health if one takes at face
value the data very recently reported by OPERA/CNGS11 data~\cite{opera}.
This is evident from the higher-significance OPERA result here reported in Eq.~(\ref{mammamia}),
but it is important for our purposes to establish that
one can conclude that  non-tachyonic special relativity is disfavored
even only using
the lower-significance
OPERA data (but with sharper information on the energy of the neutrinos) which we are using.

\noindent
To render this claim fully robust we first notice that the reduced $\chi^2$ of a fit
of non-tachyonic-special-relativity case on all the data in our Fig.~2
is an unimpressive 2.51 [df 11].
In light of the concerns expressed above for the sensitivity of results on
the large bias correction introduced in Ref.~\cite{preminos1979}
we also checked if perhaps the non-tachyonic-special-relativity case
could be rescued by
allowing the estimate of the bias $b_{1979}$ given in Ref.~\cite{preminos1979}
to vary within a 3 standard deviations range (standard deviation of $b_{1979}$ 
also estimated in Ref.~\cite{preminos1979}, as quoted above).
But by varying the bias parameter $b_{1979}$ accordingly 
we found that the reduced $\chi^2$ of the
fit can only get worse.
So we conclude that taking at face value the OPERA results and including them
in such test of hypothesis would clearly disfavor
standard non-tachyonic special relativity.

\newpage

$~$

\newpage

\subsection{Superluminal but not a tachyon}
Clearly, if taken at face value, the data presently available point toward a superluminal $\mu$ neutrino.
One could actually argue that all three results OPERA/CNGS11, MINOS/FERMILAB07 and FERMILAB79
individually favor (some more some less significantly) a $\mu$ neutrino with speeds
greater than $c$. There is a common tendency to associate the concept of
a ``superluminal particle" (speed greater than the speed-of-light scale $c$)
to a special-relativistic tachyon (particle governed by special relativity, but with imaginary mass).
This is evidently not a correct association (and readers unfamiliar with the subject may
use the illustrative examples of models here considered as guidance).
But nonetheless we find appropriate to first test the hypothesis of the $\mu$ neutrino
as a special-relativistic tachyon, as described by Eq.~(\ref{tachy}).

In Fig.~3 we show the fit of the special-relativistic-tachyon hypothesis
on the data already summarized in our Fig.~2. We computed the reduced $\chi^2$
of this fit and found a discouraging~2.26~[df 10].
Also for this special-relativistic-tachyon hypothesis we then allowed the
bias $b_{1979}$ to vary within its 3-standard-deviation range,
but could only find a negligible improvement in the reduced $\chi^2$
(2.25) in correspondence of a best-fit value for  ${\cal M}^2$
of ${\cal M}^2 = 1.13 \cdot  10^{-3}~ \text{GeV}^2$.
From the high values of reduced $\chi^2$ we conclude that 
the  special-relativistic-tachyon hypothesis is disfavored\footnote{Of course, it is a peculiar exercise to constrain
a special-relativistic tachyon hypothesis using the
more benign high-energy features rather than
the pathological implications at lower energies.
However, for reasons that will be stressed in the next section,
there is some added value for us to constrain
the special-relativistic tachyon with data at
energies between a few GeVs and 200 GeVs.}, even just
using data on $\mu$ neutrinos in the energy range from 3 to 200 GeV.

Let us then warm up to the idea of a superluminal particle
 without imaginary mass, by considering the simplest option of the Coleman-Glashow scenario
of Eq.~(\ref{colgla}).
Fig.~4 shows the result of a fit of the Coleman-Glashow parameter
on the OPERA+FERMILAB07+FERMILAB79 data already shown in our Fig.~2.
The result is satisfactory, as implied by the reduced $\chi^2$ of the fit
which we computed to be 1.26 [df 10].
Moreover, 
considering again values of $b_{1979}$ within its 3-standard-deviation range,
we found even lower values of  reduced $\chi^2$ for the fit based on
the Coleman-Glashow scenario,
including a  case with reduced $\chi^2$ of 0.70 in correspondence of
a best-fit value of the Coleman-Glashow $\delta$ parameter of $\delta = 2.6 \cdot  10^{-5} $.\\
So the Coleman-Glashow picture passes our test rather comfortably.

\begin{figure}[htbp!]
\includegraphics[width=\columnwidth]{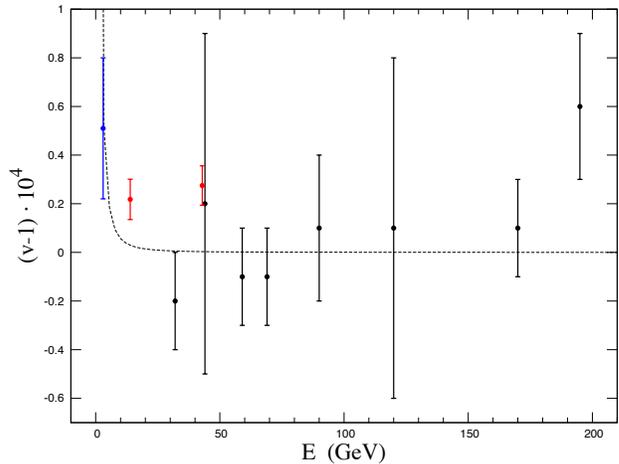}
\caption{Fit with the special-relativistic-tachyon hypothesis.}
\label{figbla}
\end{figure}

\begin{figure}[htbp!]
\includegraphics[width=\columnwidth]{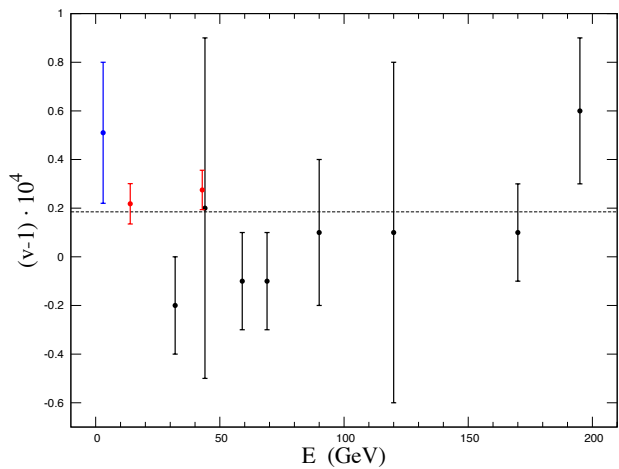}
\caption{Fit with the Coleman-Glashow hypothesis.}
\label{balbla}
\end{figure}

\newpage

$~$

\newpage

\subsection{The DSR-compatible cases}
As mentioned, the Coleman-Glashow scenario has only been studied and known to
produce acceptable physics as a scenario for a full breakdown of
special-relativistic Lorentz symmetry. Taking at face value the
available data the only special-relativistic superluminal option, the tachyon, is ``ruled out".
From the previous subsection we do have a viable candidate, the Coleman-Glashow case but
requires a preferred frame
(an ``ether frame"). Next let us explore another question: if one takes at face value
the presently-available data is it then automatic that one is forced to violate
the Relativity Principle and formulate the theory in an ``ether frame"?

We here explore this by considering the DSR-compatible cases
of Eq.~(\ref{dsr}) and  of Eq.~(\ref{dsrquad})
for which it is established that the modification of the speed law
can be implemented relativistically\footnote{For
completeness we note here (even though it is irrelevant for the
narrow scopes of the present exploratory study) that just like deforming the Galilean boosts into Lorentz boosts requires the introduction of relativity of simultaneity, we recently
understood~\cite{bob,leeINERTIALlimit,principle,grf2nd} that in turn deforming Lorentz boosts
 into DSR-Lorentz boosts requires the introduction of relativity of locality.}.

In Figs.~5 and 6 we show the results of fitting respectively
the case of Eq.~(\ref{dsr}) and  the case of Eq.~(\ref{dsrquad})
on the OPERA+FERMILAB07+FERMILAB79 data already shown in our Fig.~2.
The results are not encouraging: those fits come with a
 reduced $\chi^2$ of 2.01 [df 10] for the case of Eq.~(\ref{dsr}),
 in Fig.~5, and of 2.35 [df 10] for the case of Eq.~(\ref{dsrquad}).

For the DSR-compatible quadratic case of Eq.~(\ref{dsrquad})
the outlook does not improve much even allowing again for varying the bias $b_{1979}$
 within its 3-standard-deviation range:
the best reduced $\chi^2$ we find following this procedure is
is still of 2.06 (and would best-fit the parameter $\ell_2^2$ to the value $\ell_2^2 = 1.8 \cdot  10^{-9}
~ \text{GeV}^{-2}$).
We find this value of reduced $\chi^2$ still not encouraging, and we therefore conclude
that already with data available in the range from 3GeV to 200GeV
the the DSR-compatible quadratic case of Eq.~(\ref{dsrquad})
appears to be disfavored.

For the DSR-compatible linear case of Eq.~(\ref{dsr})
the outlook appears to be more encouraging.
By allowing  the bias $b_{1979}$ to vary
 within its 3-standard-deviation range
 one finds a sizable region with values of reduced $\chi^2$ close to 1,
 including a case with reduced $\chi^2$
of 1.10 where the fitting value of the parameter $\ell_1$ is $\ell_1 = 3.96 \cdot  10^{-7}
~ \text{GeV}^{-1}$.
We shall therefore consider the
 DSR-compatible linear case of Eq.~(\ref{dsr})
as a plausible description of the
 OPERA+FERMILAB07+FERMILAB79
 on the speed of $\mu$ neutrinos.

\begin{figure}[htbp!]
\includegraphics[width=\columnwidth]{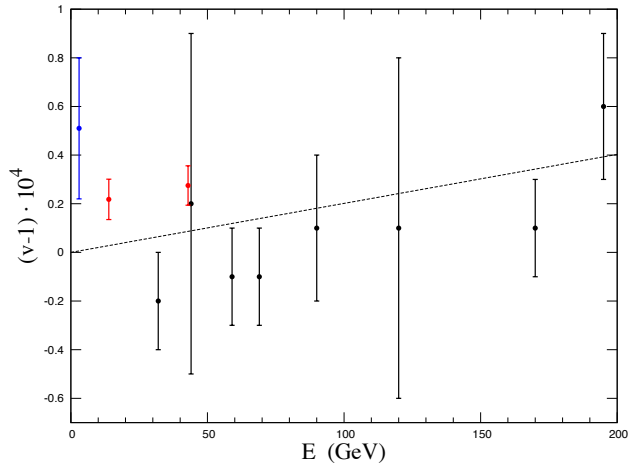}
\caption{Fit with the DSR-compatible linear case.}
\label{figbla}
\end{figure}

\begin{figure}[htbp!]
\includegraphics[width=\columnwidth]{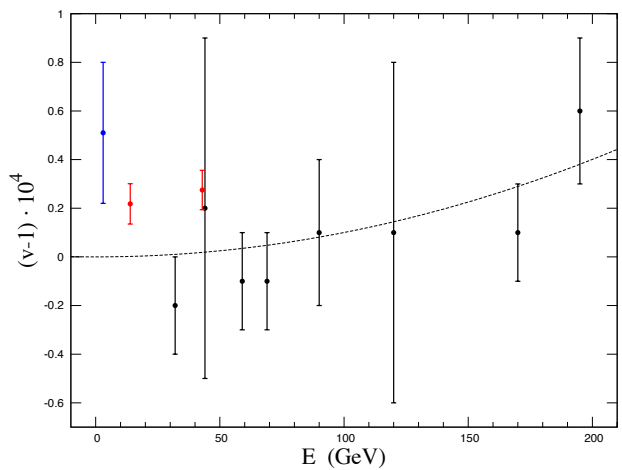}
\caption{Fit with the DSR-compatible quadratic case.}
\label{balbla}
\end{figure}

\newpage

\section{SN1987a, other neutrino species and large extra dimensions}
We have focused so far on a range of energies which is considerably wider than
the OPERA range, but still very narrow in absolute terms.
And we focused on data which apply (or can be interpreted as applying)
exclusively to $\mu$ neutrinos. We shall soon argue that there are some advantages to this
approach in a situation such as the one raised by the data recently
reported by OPERA.

But before we do that let us instead widen our horizons, considering other neutrino species
and other ranges of energy. From this perspective one should immediately consider
the observations of neutrinos from the supernova 1987a (see, {\it e.g.},
Refs.~\cite{longo,stodolsky,agliottoDIflavio}).
It is useful from this perspective to consider the Coleman-Glashow picture, which did very
well in our peculiar test on the OPERA+FERMILAB07+FERMILAB79 data, and also
the linear DSR-compatible case of Eq.~(\ref{dsr}), which in our analysis produced a plausible
description of  OPERA+FERMILAB07+FERMILAB79 data.

This happens to be also a good combination of cases on which to look at the SN1987a story,
since it allows to illustrate the difference between an energy-independent modification
of the speed law and an energy-dependent one.

Let us start with the energy-independent Coleman-Glashow case. If the same considerations
we applied above to $\mu$ neutrinos in the range of 3 to 200 GeVs, are applied universally to
neutrinos of any type and any
energy then the Coleman-Glashow picture, which provided a nice fit
of $\mu$-neutrino data between 3 GeV and 200 GeV, is immediately ruled out
by the bound $|v-1|<2 \cdot 10^{-9}$
on neutrino speeds that one robustly infers~\cite{longo,stodolsky} from
the observations of neutrinos from the supernova 1987a.
In fact, our good fit to data in the 3-200GeV range with the Coleman-Glashow picture
gave $v-1 = \delta \sim 2.6 \cdot 10^{-5}$, which in the energy-independent
Coleman-Glashow picture should be applicable to neutrinos of any energy,
including the ones of SN1987a.

The situation is not equally disastrous for scenarios where the departures
from special relativity increase with energy, such as
the DSR-compatible case of Eq.~(\ref{dsr}). These scenarios  predict effects
for lower-energy SN1987a neutrinos which are of course weaker than the ones they
predict for GeV neutrinos. So in principle one could find agreement between a larger
effect at GeV energies and a smaller effect for SN1987a neutrinos.
But the behavior must be very steep and one of the striking findings of the analysis
we reported in the previous section is that, if one empowers the analysis with all
the data available between 3GeV and 200GeV it emerges that the data situation
favors energy independence or at most a
softly increasing dependence on energy, not a very steep energy
dependence such as needed for matching the large difference in magnitude between
the feature reported by OPERA and the SN1987a bound.

And we should stress that looking at the prudent characterization of the
bound obtainable from SN1987a neutrinos given, {\it e.g.}, in Refs.~\cite{longo,stodolsky}
one can very significantly underestimate how steep the dependence on energy
would have to be in order to find agreement between
the feature reported by OPERA and the SN1987a bound.
The bound in  Refs.~\cite{longo,stodolsky}
very prudently took as reference the overall differences in times
of arrival of SN1987a neutrinos and photons, but for a scenario with energy-dependent speed
of neutrinos a much more severe constraint is obtained on the basis of the fact that
 SN1987a neutrinos of different energies reached the same detectors within a few seconds.

Let us illustrate this issue considering a generic power-law energy dependence
\begin{equation}
v - 1 = \left ( \frac{E}{M_*} \right )^\alpha
\end{equation}
for some scale $M_*$ and some power $\alpha$.
If one imposes on this {\it ansatz} the constraint that
Kamiokande observed
SN1987a neutrinos 7.5MeV and a 35MeV within 13 seconds of each other
and if one also imposes
$v-1 \sim 2 \cdot 10^{-5}$ at 20 GeV (as a rough characterization
of the result reported by OPERA)
the conclusion is that all  $\alpha < 2.5$ appear to be disfavored.

 In a companion paper now in preparation~\cite{companion} we shall
establish
 that no simple functional dependence on energy can successfully satisfy
 both the demands of our approach using data from 3 to 200 GeV on $\mu$ neutrinos
 and the constraints contained in the SN1987a story.

 We feel that if the OPERA result is taken at face value then its description cannot
 be a simple mechanism based on a universal energy-speed relation with familiarly smooth functional
 form. We expect it to be necessary to advocate strong neutrino-flavor dependence of the effect
 and/or some mechanism that switches on the superluminal behavior
 only at energies higher that the SN1987a energies.
 And while such peculiar specifications of models are evidently unappealing, it should be noticed that
 a strong flavor dependence of Lorentz-violation
 effects is for example not foreign to some frontier area
 of research, such as spacetime noncommutativity and other areas of quantum-gravity research.
 Even more noteworthy is the status of onset scales within the large-extra-dimension literature.
 As mentioned studies from more than a decade ago~\cite{ammosovvolkov,volkov}
suggested that large extra dimensions should lead to superluminal behavior of neutrinos
with an onset scale, below which the effect would be completely switched off.
 The fact that such theoretical speculations where put forward well before
 the recent  interest in superluminal neutrinos
 must be viewed  as an element of compellingness.
 And more recently (but still much before the OPERA study) other large-extra-dimension
investigations have reported the possibility of superluminal behaviour
for neutrinos, confined to specific ranges of energy~\cite{DentPas,HollePas}.

It is in light of these considerations that we expect that ultimately the main arena
for scrutinizing the intriguing OPERA result will have to be the context
of studies of the velocity of $\mu$ neutrinos in the energy range between a few GeVs
and, say, 200 or 300 GeVs.
And it is in preparation for this task that we set up in the previous section
the richest analysis of this sort that could be done on presently available data.

\section{Closing remarks}

We have shown that combining all data presently available
for $\mu$ neutrinos in the energy range from 3 GeV to
200 GeV one achieves tangible discriminating power
for candidate models of a superluminal $\mu$ neutrino.

And we have argued, at the end of the previous section,
that ultimately the main arena
for scrutinizing
the intriguing OPERA result will have to be the
context
of studies of the velocity of $\mu$ neutrinos
in the energy range between a few GeVs
and, say, 200
or 300 GeVs.

We feel that a priority for future experiments
 should be
involving the $\mu$ neutrinos of the highest
energies manageable, since this would further empower
the strategy of analysis we here proposed.

Evidently also relying on experiments with significantly
different baselines would be a major asset for the
discrimination of a propagation effect with
respect to several candidate sources of systematic bias
for velocity measurements.

\bigskip

\bigskip

\bigskip

\bigskip
{\it  GG thanks for the hospitality the APC - Universite Paris 7, where part of this work
was carried out.}


\newpage

\begin{thebibliography}{50}

\bibitem{opera}
T.~Adam {\it et al} [OPERA collaboration],
arXiv:1109.4897v1

\bibitem{minos2007}
  P.~Adamson {\it et al.} [ MINOS Collaboration ],
  Phys.\ Rev.\  { D76}, 072005 (2007).
  [arXiv:0706.0437 [hep-ex]].

\bibitem{preminos1979}
G.~R.~Kalbfleisch, N.~Baggett, E.~C.~Fowler and J.~Alspector, Phys.\ Rev.\ Lett. { 43}, 1361 (1979)

\bibitem{colgla1}
  S.~R.~Coleman, S.~L.~Glashow,
  Phys.\ Lett.\  { B405}, (1997) 249-252 .
  [hep-ph/9703240].

\bibitem{colgla2}
  S.~R.~Coleman, S.~L.~Glashow,
  Phys.\ Rev.\  { D59 } (1999)  116008.
  [hep-ph/9812418].

\bibitem{dsrnature} G. Amelino-Camelia,
arXiv:gr-qc/0207049,
Nature 418 (2002) 34-35.

\bibitem{gacdsrIJMPD2002} G.~Amelino-Camelia,
arXiv:gr-qc/0012051, { Int.~J.~Mod.~Phys.} D11 (2002)  35.

\bibitem{gacdsrPLB2001}
G.~Amelino-Camelia,
arXiv:hep-th/0012238,
Phys. Lett. B  510 (2001) 255.

\bibitem{kowadsr} J.~Kowalski-Glikman,
arXiv:hep-th/0102098, Phys.~Lett.~A 286 (2001) 391.

\bibitem{leedsrPRL}
J.~Magueijo and L.~Smolin,
arXiv:hep-th/0112090,
Phys. Rev. Lett. 88 (2002) 190403.

\bibitem{grbgac} G. Amelino-Camelia, J. Ellis, N.E. Mavromatos, D.V. Nanopoulos and S. Sarkar,
arXiv:astro-ph/9712103,
Nature 393 (1998) 763-765.

\bibitem{gampul}
R.~Gambini and J.~Pullin,
{Phys.~Rev.}~{D59} (1999) 124021.

\bibitem{mexweave} J.~Alfaro, H.A.~Morales-Tecotl and L.F.~Urrutia,
gr-qc/9909079,
Phys.~Rev.~Lett.~84 (2000) 2318.

\bibitem{leeQGLIV}
L.~Smolin,
Lect.~Notes Phys.~669 (2005) 363.

\bibitem{ammosovvolkov}
V.~Ammosov, G.~Volkov,
arXiv:hep-ph/0008032

\bibitem{volkov}
G.~G.~Volkov,
hep-ph/0607334,
Annales Fond.~Broglie 31 (2006) 227.

\bibitem{strumia}
A. de Gouvea, Gian F. Giudice, A. Strumia, K. Tobe,
Nucl.Phys.B623 (2002) 395,
arXiv:hep-ph/0107156

\bibitem{DentPas}
J.~Dent, H.~P\"as, S.~Pakvasa, T.~J.~Weiler,
arXiv:0710.2524

\bibitem{HollePas}
S.~Hollenberg, H.~P\"as
arXiv:0904.2167

\bibitem{companion}
G.~Amelino-Camelia, G.~Gubitosi, P.~Lipari, N.~Loret, F.~Mercati and G.~Rosati,
in preparation

\bibitem{sme1} V.A. Kostelecky and C.D. Lane,
Phys.~Rev.~D60 (1999) 116010.

\bibitem{smerev}
V.A.~Kostelecky and N.~Russell, arXiv:0801.0287,
Rev.Mod.Phys. 83 (2011) 11

\bibitem{ellisMAGIC}
J. Albert et al. [MAGIC Collaboration] and J.R. Ellis, N.E. Mavromatos, D.V. Nanopoulos, A.S. Sakharov and E.K.G. Sarkisyan,
Phys. Lett. B 668 (2008) 253.


\bibitem{unoEdue}
G. Amelino-Camelia and L. Smolin,
arXiv:0906.3731 [astro-ph.HE],
Phys. Rev. {D80} (2009) 084017.

\bibitem{Ellis:2008fc}
   J.~R.~Ellis, N.~Harries, A.~Meregaglia, A.~Rubbia, A.~Sakharov,
   Phys.\ Rev.\  {D78 } (2008)  033013.
   [arXiv:0805.0253 [hep-ph]].

\bibitem{bob}
  G.~Amelino-Camelia, M.~Matassa, F.~Mercati and G.~Rosati,
  arXiv:1006.2126,
  Phys.~Rev.~Lett.~{106}, 071301 (2011).

\bibitem{leeINERTIALlimit}
L.~Smolin, arXiv:1007.0718.

\bibitem{principle}
  G.~Amelino-Camelia, L.~Freidel, J.~Kowalski-Glikman
L.~Smolin,
 arXiv:1101.0931.

\bibitem{grf2nd}   G.~Amelino-Camelia, L.~Freidel, J.~Kowalski-Glikman
L.~Smolin,
 arXiv:1106.0313, Gen.~Relativ.~Gravit.~43 (2011) 2547.

\bibitem{stodolsky}
L. Stodolsky
Pys. Lett. B {201} pp. 353, (1988)

\bibitem{longo} M.~J.~Longo,
 Phys.\ Rev.\ D {36}, 3276 (1987).

\bibitem{agliottoDIflavio} P.~Galeotti and G.~Pizzella, arXiv:0706.2235
(presented at the XII-th International Workshop on Neutrino Telescope, Venice, 6-9 March 2007)

\end{thebibliography}
\end{document}